\documentclass[pra,onecolumn,aps]{revtex4}

\usepackage{amssymb}
\usepackage{amsmath}
\usepackage[pdftex]{graphicx}

\begin{document}

\title{Shoot regeneration seedlings from transverse thin cell layer explants 
excised from cotyledon, petiole, hypocotyl of \textit{Brassica juncea} L. 
Czern. in the presence of CdCl$_2$}

\author{ Michel AOUN\S}\email{ michel.aoun@univ-brest.fr}
\author{Jean-Yves CABON\P, Annick HOURMANT\S}
 \affiliation{\S Laboratoire de Biotechnologie et Physiologie V\'eg\'etales, 
Universit\'e de Bretagne Occidentale (Brest - France). \\
\P Laboratoire de Chimie, Electrochimie Mol\'eculaire et Chimie Analytique - 
UMR 6521, Universit\'e de Bretagne Occidentale (Brest - France).}

\begin{abstract}
The present work describes a novel neoformation process of \textit{Brassica 
juncea} (L.) Czern. seedlings from transverse thin cell layers in the presence 
of CdCl$_2$. In order to investigate the regeneration ability of this crop, the 
effect of CdCl$_2$ on shoot regeneration (frequency of regeneration and bud 
number per tTCL) was examined. The tTCL explants were excised from cotyledon, 
petiole and hypocotyl of 7 day-old \textit{B. juncea} seedlings and cultivated 
on a solid basal MS medium supplemented with $\alpha $-naphtalenacetic acid 
(NAA : 3.22 $\mu$M), 6-benzylamino-purine (BAP : 26.6 $\mu$M), sucrose (2 \%, 
w/v), silver nitrate (AgNO3 : 10 $\mu$M) and various concentrations of CdCl$_2$ 
(0 - 250 $\mu$M).    \\
A concentration of CdCl$_2$ of 5 $\mu$M is enough to reduce 
significantly the percentage of regenerated tTCL from 95-100 \% to 77-86 \% for 
all organs tested. In addition, 5 $\mu$M of CdCl$_2$ reduces the bud number 
from 4.5 to 2.98 and 3.2 to 2.02 of hypocotyl and petiole tTCLs, but not for 
cotyledon tTCLs. Besides, 250 $\mu$M of CdCl$_2$ is lethal for all tTCLs 
whatever the organ, and 200 $\mu$M is lethal only for cotyledon and petiole 
tTCLs but not for hypocotyl explants which had 3.6 \% of frequency of shoot 
regeneration. Plantlets regenerated from all shoots, whatever the treatment, 
developed and flowered normally 6 weeks after the transfer to pots. The 
regenerated plants were fertile and identical to source plants.

\end{abstract}

\maketitle


\textbf{Abbreviations}   BAP  6-benzylamino-purine; NAA  $\alpha 
$-naphtalenacetic acid; 
MS  Murashige and Skoog's medium (1962); PGRs  Plant growth regulators; tTCL(s) 
Transverse thin cell layer(s)  \\

\section{Introduction}
\textit{Brassica} oilseed crops, \textit{Brassica juncea}, \textit{B. napus} 
and \textit{B. rapa}, cover more than 11 million hectares of the world's 
agricultural land and provide over 8\% of the major oil when grown under a 
variety of climatic conditions (Downey, 1990). The annual production of 
\textit{Brassica juncea} in India reaches alone five million tons (FAO, 2003).   \\

In the recent past decades, \textit{Brassica juncea} has drawn the attention of 
researchers because of its high biomasse production with added economical value 
and its high capacity to translocate and cumulate many metals and metalloids as 
As, Cd, Cu, Pb, Se, and Zn from polluted soils and therefore, can be considered 
as a good candidate in phytoremediation processes (Kumar \textit{et al.} 1995; 
Salt \textit{et al.} 1995).   \\

Research focused mainly on the transgenic approach in order to increase the 
capacity of \textit{Brassica juncea} or other plants to cumulate high metal 
levels (Lasat, 2002; Pilon-Smits et Pilon, 2002). However, to the best of our 
knowledge, there is no report on the \textit{in vitro} selection of 
\textit{Brassica juncea} using tTCLs excised from cotyledons, petioles and 
hypocotyls of young plants in the presence of Cd$^{2+}$ which can result into a 
novel approach to obtain a regenerant cumulating more Cd than the source plants 
and use in many phytoremediation processes.   \\

Cadmium like other metals such as Al, Cu, Pb and Zn is toxic for humans, 
animals and plants and is a widespread contaminant with a long biological 
half-life (Wagner, 1993; Das \textit{et al.} 1997). Cd, that is easily absorbed 
by plants, has a major toxic effect on their photosynthesis and their growth 
(Prasad, 1995; Briat and Lebrun, 1999). The growth inhibition has been mainly 
studied on whole plants originating from bean seeds (Poschenrieder \textit{et 
al}. 1989), willow, poplar (Lunackova \textit{et al}. 2003; Cosio \textit{et 
al}. 2005), rice (Aina \textit{et al.} 2007), sunflower (Groppa \textit{et al.} 
2007), and some \textit{Brassica} species such as \textit{Brassica napus} 
(Larsson \textit{et al.} 1998) and \textit{Brassica juncea} (Haag-Kerwer 
\textit{et al.} 1999). However, to the best of our knowledge, there is no 
report on the effect of Cd on plant regeneration by \textit{in vitro} culture.   \\

In the present paper, the effect of CdCl$_2$ on shoot regeneration (frequency 
of shoot regeneration and number of bud per explant) from tTCLs is investigated 
in Indian mustard.    \\

\section{Material and methods}

\textbf{ Plant material}\\

\textit{Brassica juncea} AB79/1 was used to evaluate shoot regeneration in 
presence or absence of CdCl$_2$. This cultivar is pure spring line, genetically 
fixed and was obtained by autofertilization.   \\

\subsection{ Culture condition and regeneration of plants}

Seeds of \textit{Brassica juncea} were decontaminated in 70\% ethanol for 30 
sec, followed by immersion in calcium hypochlorite (5\%, w/v) added with two 
drops of Tween-20 for 10 min. The seeds were rinsed twice for 5 min with 
sterile water upon sterilization and sown in test tubes on MS medium containing 
20 g.l$^{-1}$ of sucrose and solidified with agar at 6 g.l$^{-1}$ (Kalys, HP 
696). They were incubated later on under a photoperiod of 12h (60 $\mu$mol 
photon.m$^{-2}$s$^{-1}$) provided by cool white fluorescent lamps, with a 
22/20${}^\circ$C thermoperiod (light/dark).   \\

tTCLs (400-500 $\mu$m) were excised from cotyledons, petioles and hypocotyls of 
7 day-old \textit{Brassica juncea} seedlings and cultivated in a Petri dishes 
containing MS medium (25 ml) (15 tTCLs per Petri dish). Cotyledon, petiole and 
hypocotyl tTCLs were cultivated on MS medium (comprising macronutriments, 
micronutriments and vitamins of Murashige and Skoog, 1962) supplemented with 
BAP (26.6 $\mu$M), NAA (3.22 $\mu$M), Sucrose (20 g.l$^{-1}$), AgNO3 (10 
$\mu$M) and various concentrations of CdCl$_2$ (0-250 $\mu$M). All media were 
solidified with agar (0.6 \%, w/v), adjusted to pH 5.8 by 0.1 N NaOH and 
sterilized by autoclaving at 120${}^\circ$C for 20 min.   \\

All cultures were incubated in the same conditions as previously described. The 
number of explants with shoot buds was recorded after 4 weeks culture and the 
number of adventitious shoots per tTCL were counted. After 4 weeks, shoots were 
separated and transferred to test tubes containing MS medium (10 ml) without 
any PGRs to induce rooting. The small plantlets were transferred to pots 
containing sterile vermiculite (EFISOL, VERMEX M) in a naturally-lighted 
greenhouse, watered daily and fertilized with half strength Hoagland solution 
(Hoagland and Arnon, 1950). All transfered plants flowered normally and were 
identical to source plants.   \\

\section{Measurement of Cd in the neoformed shoots}

The determination of cadmium concentrations in the different digested solutions 
was conducted by electrothermal atomic absorption spectrometry. A Perkin-Elmer 
SIMAA 6100 working in the single element monochromator mode was used for all 
atomic absorption measurements. At harvest (27 days after the initiation of 
tTCL culture), shoots were weighed and then oven-dried for 4 days at 
80${}^\circ$C. For the preparation of all solutions, high- purity water from a 
MilliQ-system (Millipore, Milford, MA, USA) was used. Sample aliquots of 
approximately 200 mg were transferred into the Teflon vessels. After addition 
of acid mixture : nitric acid, hydrogen peroxide and hydrofluoric acid to the 
powders in the ratios (4/3/1, v/v/v), the vessels were closed and exposed to 
microwaves digestion as described in detail elsewhere (Weiss \textit{et al.} 
1999).   \\

\subsection{ Data analysis}

The frequency of shoot regeneration and the number of shoots par tTCL was 
recorded from 5 replicates, each with 15 tTCLs per Petri dish. Each experiment 
was repeated 3 times with 3 independent runs. The values were compared by 
analysis of variance (ANOVA) and the differences among means (5\% level of 
significance) were tested by the LSD test using StatGraphics Plus 5.1.

\subsection{Results}

Shoot regeneration occured from tTCL explants 10 days after the tTCL initiation 
culture when the tTCLs were cultured in the absence of CdCl$_2$ whatever the 
source explants, but slightly later when this metal was applied. This 
inhibition of tTCL initiation culture was function of the metal concentration 
levels (Table 1). The largest number of occurence was obtained at day-27 for 
all organ tested and whatever the final concentration of CdCl$_2$; the number 
of occurrence unchanged thereafter. When the optimal conditions of culture were 
present, we obtained the maximal frequency of tTCL forming buds (100, 98.87 and 
93.78 \% for cotyledon, petiole and hypocotyl tTCLs respectively) (Table 1). 
Optimal conditions of culture were obtained when MS solidified medium was 
supplemented with NAA (3.22 $\mu$M), BAP (26.6 $\mu$M), sucrose (2\%, w/v) and 
AgNO3 (10 $\mu$M) (unpublished data obtained in our laboratory). Furthemore, in 
the same conditions, the origin of tTCL explants affected the number of buds 
per tTCL. Indeed, with the hypocotyl tTCLs, we obtained the best number of 
buds, followed by petiole then cotyledon tTCLs (3.20 and 2.40 respectively).   \\

Besides, 5 $\mu$M of CdCl$_2$ was enough to reduce significantly, the rate of 
shoot regeneration and the number of buds per tTCL explant except the number of 
buds per cotyledon tTCLs which stayed unchanged (Table 1). When the 
concentration of CdCl$_2$ increased, the frequency of shoot regeneration and 
the number of buds per explant decreased (Table 1). High Cd concentration (250 
$\mu$M) in the culture medium was lethal for callus culture and blocked shoot 
regeneration. Depending on source explants, hypocotyl tTCLs were more tolerant 
to CdCl$_2$ than cotyledon and petiole tTCLs. Our results show that only 
hypocotyl tTCLs were able to regenerate in the presence of 200 $\mu$M of 
CdCl$_2$ (3.56\% and 0.07 bud per explant) (Table 1).   \\

In addition, Cd contents were estimated in shoot buds. We observed that the 
latter increased in shoots with the increase of metal concentrations in the 
culture medium (Table 2). The highest content was observed with 150 $\mu$M of 
CdCl$_2$. However, there is no difference between organs for Cd cumulation in 
neoformed buds (Table 2). Regenerated plantlets in the presence or the absence 
of CdCl$_2$, whatever its concentration, were able to grow and flower normally 
6 weeks after their transfer in the greenhouse (Figures 1 and 2).   \\

\subsection{Discussion and conclusion}

Based on the efficiency of the thin cell layer technology for the propagation 
of various plant species, even for the more recalcitrant ones, this study was 
undertaken to evaluate the effect of CdCl$_2$ on the frequency of shoot 
regeneration and bud numbers per tTCLs in \textit{B. juncea} L. Czern. from 
tTCL explants excised transversally from young axenic plants.   \\

In our experiments, we showed a maximal frequency of regeneration in MS basal 
medium supplemented with NAA (3.22 $\mu$M), BAP (26.6 $\mu$M), sucrose (2\%, 
w/v) and AgNO3 (10 $\mu$M). There is no significant difference in the frequency 
of shoot regeneration between the 3 organs tested. However, we observed a 
significant difference of bud numbers per explants between organs (Table 1). 
Indeed, for all organs, hypocotyl tTCLs exhibited the highest bud number per 
explant, followed by petiole and cotyledon tTCLs respectively. The effect of 
explant source on shoot regeneration was previously reported in traditional 
leaf and cotyledon explants in \textit{B. juncea} var. Tsatsai (Guo \textit{et 
al.} 2005), longitudinal thin cell layers (lTCLs) from stems (Klimaszewska and 
Keller, 1985; Nhut \textit{et al.} 2003) and cotyledon explants (Ono \textit{et 
al.} 1994) and more recently, in hypocotyl and petiol transversal thin cell 
layers (tTCLs) in \textit{B. napus} L. (Ben Ghnaya \textit{et al.} 2008).   \\

The presence of CdCl$_2$ in the culture medium showed a negative impact on, 
both, the frequency of shoot regeneration and the number of buds per tTCL, even 
when the concentration of CdCl$_2$ was low (e.g. 5 $\mu$M). High concentrations 
of CdCl$_2$ (200 and 250 $\mu$M) were critical for the survival and the 
neoformation process of tTCL explants except hypocotyl tTCLs which were able to 
regenerate at 200 $\mu$M of CdCl$_2$ (Table 1). Therefore, these results showed 
source explant variation in response to this metal. To the best of our 
knowledge, there is no report on this explant variation response to CdCl$_2$ in 
\textit{B. juncea} L. Czern.   \\

It was known that cadmium is not essential for plant growth and development, 
and it was shown to inhibit the growth of many plant species such as bean 
(Poschenrieder \textit{et al}. 1989), willow, poplar (Lunackova \textit{et al}. 
2003; Cosio \textit{et al}. 2005), rice (Aina \textit{et al.} 2007), sunflower 
(Groppa \textit{et al.} 2007), and some \textit{Brassica} species such as 
\textit{Brassica napus} (Larsson \textit{et al.} 1998) and \textit{Brassica 
juncea} (Haag-Kerwer \textit{et al.} 1999). However, all these studies on Cd 
toxicity were performed in plants which came directly from seeds. We believe, 
as well as Gladkov (2007), are the first authors who obtained regenerated 
plants in the presence of Cd. Indeed, Gladkov (2007) showed that, the 
inhibitory effect of Cd on \textit{Agrostis stolonifera} and red fescue 
(\textit{Festuca rubra}) callus cultures, was observed at Cd$^{2+}$ 
concentration of 7 mg.l$^{-1}$; at 20 and 30 mg.l$^{-1}$, considerable 
proportions of callus cells darkened and died, and a concentration of 60 
mg.l$^{-1}$ was lethal to the cultures. He obtained some regenrated plants (30 
from \textit{A. stolonifera} and 4 from \textit{F. rubra}).   \\

In our study, we observed that neoformed buds were able to accumulate Cd$^{2+}$ 
in their tissues. This accumulation was function of Cd concentration. 
Therefore, an increase of Cd amounts is correlated to an increase of 
concentration of Cd in the culture medium. These results suggest a 
translocation of Cd from medium to calli and then to the buds.   \\

Regenerated plantlets were able to grow and flower normally 6 weeks after their 
transfer in the greenhouse and that whatever the concentration of CdCl$_2$ and 
their identical to source plants (Figures 1 and 2).   \\

To conclude, our model of shoot regeneration in the presence of Cd should be 
considered as a novel approach to \textit{in vitro} selection of tolerant 
regenerants which should be used in different phytoremediation processes to 
test their ability in the depollution of contaminated soils and their capacity 
to cumulate heavy metals. We suggest that \textit{in vitro} selection process 
using tTCL technology in presence of heavy metals is a novel interesting 
alternative which, in the near futur, can be used as the transgenic approach 
for improving the capacity of \textit{B. juncea} or other hyperaccumulator 
plants to tolerate, cumulate and translocate more Cd or other heavy metals than 
the original wild type. \\

\vspace{1cm}

\textbf{Acknowledgement}

We thank Dr T. Guinet from Ecole Nationale d'Enseignement Sup\'erieure 
Agronomique (ENESA) de Dijon (France) for furnishing the seeds of spring line 
AB79/1 of \textit{Brassica juncea}.\\

\vspace{1cm} 

\textbf{References}

Aina, R., Labra, M., Fumagalli, P., Vannini, C., Marsoni, M., Cucchi, U., 
Bracale, M., Sgorbati, S., Citterio, S., 2007. Thiol-peptide level and 
proteomic changes in response to cadmium toxicity in \textit{Oryza sativa} L. 
roots. \textit{Environ. Exp. Bot.} 59, 381-392.

Ben Ghnaya, A., Charles, G., Branchard, M., 2008. Rapid shoot regeneration from 
thin cell layer explants excised from petioles and hypocotyls in four cultivars 
of \textit{Brassica napus} L. \textit{Plant Cell Tiss. Organ Cult.} 92, 25-30.

Briat, J.-F., Lebrun M., 1999. Plant response to metal toxicity.  \textit{Plant 
Biol. Pathol. C.R. Acad. Sci. Paris, Sciences de la vie/ Life sciences} 322. 
43-54.

Cosio, C., Vollenweider, P., Keller, C., 2005. Localization and effects of 
cadmium in leaves of a cadmium-tolerant willow (\textit{Salix viminalis }L.). 
I. Macrolocalization and phytotoxic effects of cadmium. \textit{Environ. Exp. 
Bot. }

Das, P., Samantaray, S., Rout, G.R., 1997. Studies on cadmium toxicity in 
plants : A review. \textit{Environ. Pollut.} 98, 29-36.

Downey, R.K., 1990. \textit{Brassica} oilseed breeding-achievements and 
opportunities. \textit{Plant Breed.} 60, 1165-1170.

Food and Agriculture Organization, FAO, 2003. \textit{FAO bulletin of 
statistics}, Vol. 4 No 2.

Gladkov, E., 2007. Effect of complex interaction between heavy metals on plant 
in a megalopolis. \textit{Rus. J. Ecol.} 38, 68-71.

Groppa, M.D., Ianuzzo, M.P., Tomaro, M.L., Benavides, M.P., 2007. Polyamine 
metabolism in sunflower plants under long-term cadmium or copper stress. 
\textit{Amino Acids} 32, 265-275.

Guo, D.-P., Zhu, Z.-J., Hu, X.-X., Zheng, S.-J., 2005. Effects of cytokinins 
on shoot regeneration from cotyledon and leaf segment of stem mustard 
\textit{Brassica juncea} var. Tsatsai. \textit{Plant Cell Tiss.Org.Cult.} 83, 
123-127.

Haag-Kerwer, A., Sch\"{a}fer, H.J., Heiss, S., Walter, C., Rausch, T., 1999. 
Cadmium exposure in \textit{Brassica juncea} causes a decline in transpiration 
rate and leaf expansion without effect on photosynthesis. \textit{J. Exp. Bot. 
}50, 1827-1835.

Hoagland, D.R., Arnon, D.I., 1950. The water-culture for growing plants 
without soil. \textit{Cal. Agric. Exp. Sta Cir.} 347, 1-32.

Klimaszewska, K., Keller, W.A., 1985. High frequency plant regeneration from 
thin layer explants of \textit{Brassica napus}. \textit{Plant Cell Tiss. Organ 
Cult.} 4, 183-197.

Kumar, P.B.A.N., Dushenkov, V., Motto, H., Raskin, I., 1995. Phytoextraction : 
The use of plants to remove heavy metals from soils. \textit{Environ. Sci. 
Technol.} 29, 1232-1238.

Larsson, E.H., Bornman, J.F., Asp, H., 1998. Influence of UV-radiation and 
Cd$^{2+}$ on chlorophyll fluorescence, growth and nutrient content in 
\textit{Brassica napus}. \textit{J. Exp. Bot.} 323, 1031-1039.

Lasat, M.M., 2002. Phytoextraction of toxic metals : A review of biological 
mechanisms. \textit{J. Environ. Qual.} 31, 109-120.

Lunackova, L., Sottnikova, A., Masarovicova, E., Lux, A., Stresko, V., 2003, 
Comparison of cadmium effect on willow and poplar in response to different 
cultivation conditions. \textit{Biol. Plant.} 47, 403-411.

Nhut, D.T., Teixeira da Silva, J.A., Le, B.V., Tran Thanh Van, K., 2003. Thin 
cell layer studies of vegetable, leguminous and medicinal plants, Chapter 10. 
\textit{In}: Nhut, D.T., Tran Thanh Van, K., Le, B.V., Thorpe, T. (eds) Thin 
cell layer culture system: regeneration and transformation applications. Kluwer 
Academic Publishers, Dordrecht, pp 387-426.

Ono, Y., Takahata, Y., Kaizuma, N., 1994. effect of genotype on shoot 
regeneration from cotyledonary explants of rapeseed (\textit{Brassica napus} 
L.). \textit{Plant Cell. Rep.} 14, 13-17.

Salt, D.E., Blaylock, M., Kumar, N., Dushenkov, V., Ensley, B.D., Chet, I., 
Raskin, I., 1995. Phytoremediation : A novel strategy for removal of toxic 
metals from the environment using plants. \textit{Biotech.} 13, 468-474.

Pilon-Smits, E.A.H., Pilon, M., 2002. Phytoremediation of metals using 
transgenic plants. \textit{Crit. Rev. Plant Sci. }21, 439-456.

Poschenreider, C., Guns\'e, B., Barcelo, J., 1989. Influence of cadmium on 
water relations, stomatal resistance and abscisic acid content in expanding 
bean leaves. \textit{Plant Physiol.} 90, 1365-1371.

Prasad, M.N.V., 1995. Cadmium toxicity and tolerance in vascular plants. 
\textit{Environ. Exp. Bot.} 35, 525-545.

Wagner, G.J., 1993. Accumulation of cadmium in crop plants and its 
consequences to human health. \textit{Adv. Agron.} 51, 173-212.

Weiss, D.J., Shotyk, W., Schafer, J., Loyall, U., Grollimund, E., Gloor, M., 
1999. Microwave digestion of ancient peat and lead determination by 
voltammetry.\textbf{ }\textit{Fresenius J. Anal. Chem}. 363, 300-305.

\begin{widetext}
\begin{table}[htbp]
\begin{center}
\caption{Effect of CdCl$_{2}$ on \textit{in vitro} organogenesis 
from tTCLs of cotyledons, petioles and hypocotyls.} 
\label{tab1}
\begin{tabular}{ c c c c c c c}
\hline
CdCl$_{2 }$  & \multicolumn{3}{c}\underline {Shoot regeneration frequency 
({\%})} &  
\multicolumn{3}{c}\underline { Number of buds 
per regenerating tTCL}  \\
& Cotyledons & Petioles &  Hypocotyls & Cotyledons & Petioles & Hypocotyls \\

\hline

0 &   \textbf{100}$^{a }$&  \textbf{98.87}$^{a }$&  \textbf{93.78}$^{a }$&  
\textbf{2.40}$^{c }$&  \textbf{3.20}$^{b }$&  \textbf{4.24}$^{a}$ \\

5 &  86.22$^{b }$&  80.89$^{bc }$&  77.33$^{bc }$&  2.36$^{c }$&  2.02$^{cd }$& 
 2.98$^{b}$ \\

25 &  85.78$^{b }$&  67.56$^{c }$&  67.11$^{c }$&  1.73$^{d }$&  1.34$^{d }$&  
2.47$^{c}$ \\

50 &  70.22$^{c }$&  63.56$^{c }$&  66.67$^{c }$&  0.97$^{e }$&  1.17$^{de }$&  
1.88$^{d}$ \\

100&   22.20$^{e }$&  31.11$^{d }$&  35.56$^{d }$&  0.29$^{g }$&  0.44$^{f }$&  
0.85$^{e}$ \\

150&   0.44$^{g }$&  21.30$^{e }$&  23.30$^{e }$&  0.04$^{h }$&  0.23$^{g }$&  
0.50$^{f}$ \\

200&   0$^{h }$&  0$^{h }$&  \textbf{3.56}$^{f }$&  0$^{g }$&  0$^{g }$&  
\textbf{0.07}$^{h}$ \\

250&   0$^{h }$&  0$^{h }$&  0$^{h }$&  0$^{g }$&  0$^{g }$&  0$^{g}$  \\

\hline
\end{tabular}
\end{center}
\end{table}

\end{widetext}

Data (shoot regeneration frequency and number of buds per regenerating tTCL) 
were collected after 4 weeks of culture on MS basal medium supplemented with 
NAA (3.22 $\mu $M), BAP (26.6 $\mu $M), Sucrose 2 {\%} (w/v) and AgNO$_{3}$ 
(10 $\mu $M) and various concentrations of CdCl$_{2}$ (0-250 $\mu $M).

The results were calculated from three independent experiments, each with, 
at least, five dishes with 15 tTCLs per dish. For each parameter, the values 
with different letters are significantly different at $p$ = 0.05 (LSD test).

\begin{table}[htbp]
\begin{center}
\caption{ Cadmium contents in regenerated shoots obtained from 27 day-old 
tTCLs.} 
\label{tab2}
\begin{tabular}{ c c c c}
\hline  
 CdCl$_{2 }(\mu $M) & \multicolumn{3}{c} \underline {Cd content per bud ($\mu 
$g.g$^{-1}$ D.W.)}  \\
&  Cotyledons &  Petioles&   Hypocotyls \\

\hline

0 &  0$^{e }$&  0$^{e }$&  0$^{e}$ \\

25 &  29.23$^{d }$&  30.43$^{d }$&  33.01$^{d}$ \\

50 &  76.45$^{c }$&  78.57$^{c }$&  82.84$^{c}$ \\

100 &  166.32$^{b }$&  168.45$^{b }$&  172.34$^{b}$ \\

150&   \textbf{278.87}$^{a }$&  \textbf{281.56}$^{a }$&  \textbf{285.87}$^{a}$  
\\

\hline
\end{tabular}
\end{center}
\end{table}

Data (Cd contents in regenerated shoots) were collected after 27 days of 
culture on MS basal medium supplemented with NAA (3.22 $\mu $M), BAP (26.6 
$\mu $M), Sucrose 2 {\%} (w/v), AgNO$_{3}$ (10 $\mu $M) and various 
concentrations of CdCl$_{2}$ (0-150 $\mu $M).

The results were calculated from three independent experiments, each with, 
at least, five dishes with 15 tTCLs per dish. For Cd contents, the values 
with different letters are significantly different at $p$ = 0.05 (LSD test).

\begin{figure}[htbp]
\begin{center}
\scalebox{0.8}{\includegraphics*{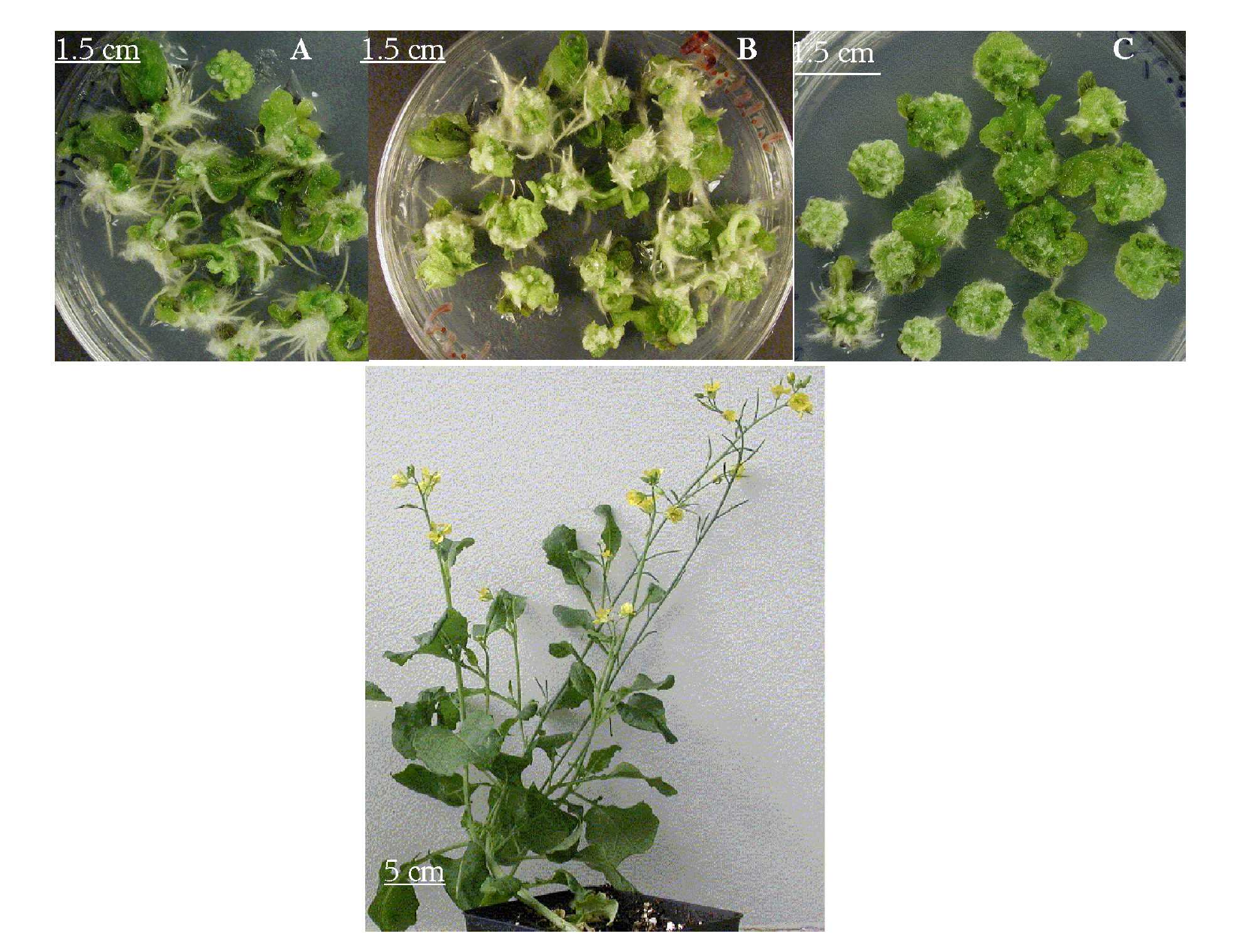}}
\end{center}
\caption{Shoot regeneration from tTCLs of Indian mustard in the absence of 
CdCl$_2$. (A)~: cotyledon tTCLs~; (B)~: petiole tTCLs~; (C)~: hypocotyl tTCLs~; 
(D)~: regenerated plants grown and flowered in the greenhouse 6 weeks after the 
transfer in pot.} 
\label{fig1}
\end{figure}

\begin{figure}[htbp]
\begin{center}
\scalebox{0.8}{\includegraphics*{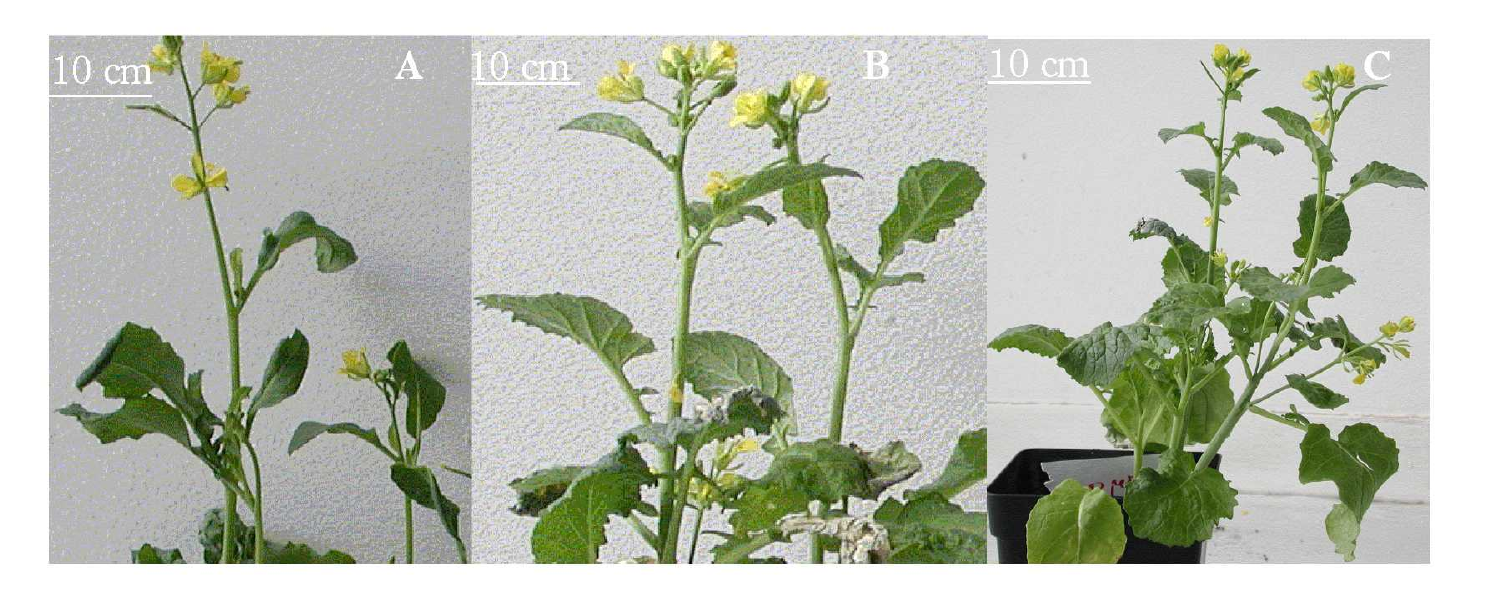}}
\end{center}
\caption{ Regenerated plants grown and flowered normally in the greenhouse 
after 6 weeks after transfer in pot. (A)~: Shoot regeneration from cotyledon 
tTCLs in the presence of 150 $\mu$M of CdCl$_2$~; (B)~: Shoot regeneration from 
petiole tTCLs in the presence of 150 $\mu$M of CdCl$_2$~; (C)~: Shoot 
regeneration from hypocotyl tTCLs in the presence of 200 $\mu$M of 
CdCl$_2$.(bar~: 10 cm)
} \label{fig2}
\end{figure}

\end{document}